\documentclass[12pt]{article}
\usepackage{graphicx}

\def\hybrid{\topmargin 0pt      \oddsidemargin 0pt
        \headheight 0pt \headsep 0pt
       \voffset-1cm
        \textwidth 6.25in       
       \textheight 9.5in       
        \marginparwidth 0.0in
        \parskip 5pt plus 1pt   \jot = 1.5ex}
\catcode`\@=11
\def\marginnote#1{}

\newcount\hour
\newcount\minute
\newtoks\amorpm
\hour=\time\divide\hour by60
\minute=\time{\multiply\hour by60 \global\advance\minute by-\hour}
\edef\standardtime{{\ifnum\hour<12 \global\amorpm={am}%
        \else\global\amorpm={pm}\advance\hour by-12 \fi
        \ifnum\hour=0 \hour=12 \fi
        \number\hour:\ifnum\minute<10 0\fi\number\minute\the\amorpm}}
\edef\militarytime{\number\hour:\ifnum\minute<10 0\fi\number\minute}

\def\draftlabel#1{{\@bsphack\if@filesw {\let\thepage\relax
   \xdef\@gtempa{\write\@auxout{\string
      \newlabel{#1}{{\@currentlabel}{\thepage}}}}}\@gtempa
   \if@nobreak \ifvmode\nobreak\fi\fi\fi\@esphack}
        \gdef\@eqnlabel{#1}}
\def\@eqnlabel{}
\def\@vacuum{}
\def\draftmarginnote#1{\marginpar{\raggedright\scriptsize\tt#1}}

\def\draftlabel#1{{\@bsphack\if@filesw {\let\thepage\relax
   \xdef\@gtempa{\write\@auxout{\string
      \newlabel{#1}{{\@currentlabel}{\thepage}}}}}\@gtempa
   \if@nobreak \ifvmode\nobreak\fi\fi\fi\@esphack}
        \gdef\@eqnlabel{#1}}
\def\@eqnlabel{}
\def\@vacuum{}
\def\draftmarginnote#1{\marginpar{\raggedright\scriptsize\tt#1}}

\def\draft{\oddsidemargin -.5truein
        \def\@oddfoot{\sl preliminary draft \hfil
        \rm\thepage\hfil\sl\today\quad\militarytime}
        \let\@evenfoot\@oddfoot \overfullrule 3pt
        \let\label=\draftlabel
        \let\marginnote=\draftmarginnote
   \def\@eqnnum{(\theequation)\rlap{\kern\marginparsep\tt\@eqnlabel}%
\global\let\@eqnlabel\@vacuum}  }

\def\numberbysection{\@addtoreset{equation}{section}
        \def\theequation{\thesection.\arabic{equation}}}

\def\underline#1{\relax\ifmmode\@@underline#1\else
        $\@@underline{\hbox{#1}}$\relax\fi}

\def\titlepage{\@restonecolfalse\if@twocolumn\@restonecoltrue\onecolumn
     \else \newpage \fi \thispagestyle{empty}\c@page\z@
        \def\thefootnote{\fnsymbol{footnote}} }

\def\endtitlepage{\if@restonecol\twocolumn \else  \fi
        \def\thefootnote{\arabic{footnote}}
        \setcounter{footnote}{0}}  
\relax

\hybrid

\newfont{\Bbb}{msbm10 scaled 1\@ptsize00}
\newfont{\Bbbb}{msbm7 scaled 1\@ptsize00}

\newcommand{\DDD}{\raise-1pt\hbox{$\mbox{\Bbbb D}$}}

\newcommand{\UUU}{\raise-1pt\hbox{$\mbox{\Bbbb U}$}}

\newcommand{\z}{\raise-1pt\hbox{$\mbox{\Bbbb Z}$}}

\def\beq{\begin{equation}}
\def\eeq{\end{equation}}
\def\p{\partial}

\newtheorem{lemma-definition}{Lemma-Definition}[section]

\begin{document}

\begin{titlepage}

\title{How Calogero-Moser particles can stick together}

\author{A. Zabrodin\thanks{Steklov Mathematical Institute of Russian Academy of Sciences,
Gubkina str. 8, Moscow, 119991, Russian Federation,
e-mail: zabrodin@itep.ru}}

\date{January 2021}
\maketitle


\begin{abstract}

We show that the configuration in the phase space of the elliptic Calogero-Moser
model when particles stick together in pairs is stable under the third Hamiltonian flow
of the model. The equations of motion for the pairs coincide with the equations of motion
for poles of elliptic solutions to the B-version of the Kadomtsev-Petviashvili equation.

\end{abstract}

\end{titlepage}


%



\section{Introduction}

This paper is a short remark on how the equations of motion for poles of
elliptic solutions to the B-version of the Kadomtsev-Petviashvili (BKP) equation
can be obtained entirely in terms of the elliptic Calogero-Moser model
\cite{Calogero71,Calogero75,Moser75,OP81}. 
Since Krichever's seminal paper \cite{Krichever80} it is known that poles of elliptic
solutions to the KP equation move as particles of the elliptic Calogero-Moser many-body
system. At the same time, solutions to the BKP equation 
\cite{DJKM83,DJKM82} are those solutions to the KP equation,
\beq\label{kp1}
3u_{yy}=\left( 4u_{t}-12u u_x -u_{xxx}\right)_x,
\eeq
for which the tau-function, i.e the entire function $\tau =\tau(x,y,t)$ such that
\beq\label{kp2}
u=\p_x^2\log \tau ,
\eeq
at $y=t_2=0$ is a full square, i.e., all its zeros are of the second order. 
Poles of the elliptic solutions are zeros of the
tau-function which, therefore, can be identified
with coordinates of the Calogero-Moser particles. 
This means that equations of motion for poles of solutions to the BKP equation,
which are elliptic functions in $t_1=x$,
derived in \cite{RZ19} should follow from restriction of the third Hamiltonian flow $t=t_3$ 
of the 
Calogero-Moser model to the subspace of the phase space in which $N=2n$ particles stick
together in pairs, i.e. their coordinates satisfy $x_{2i-1}=x_{2i}$, $i=1, \ldots , n$.
(However, one can not directly put $x_{2i-1}=x_{2i}$ in the equations of motion
because this limit is singular;
instead, one should put $x_{2i}-x_{2i-1}=O(\varepsilon)$ 
and consider the limit $\varepsilon \to 0$.)
As is easily seen, 
this configuration is immediately 
destroyed under the standard (second) Hamiltonian flow $y=t_2$ of the 
Calogero-Moser model. Nevertheless, we show that it is stable under the third Hamiltonian flow,
derive equations of motion for the pairs and show that they coincide with the equations of
motion for poles of elliptic solutions to the BKP equation obtained in \cite{RZ19}.

The elliptic $N$-body Calogero-Moser system is known to be
integrable: it has $N$ integrals of motion (Hamiltonians) in involution.
The first three are (see, e.g., \cite{UWH93})
\beq\label{ham}
\begin{array}{l}
\displaystyle{
H_1=-\sum_{i=1}^N p_i,}
\\ \\
\displaystyle{
H_2=\sum_{i=1}^N p_i^2-\sum_{i\neq j}\wp (x_i-x_j),}
\\ \\
\displaystyle{
H_3=-\sum_{i=1}^N p_i^3+3\sum_{i\neq j}p_i \, \wp (x_i-x_j),}
\end{array}
\eeq
where $\wp (x)$ is the elliptic Weierstrass $\wp$-function 
with complex periods $2\omega_1$, $2\omega_2$
such that ${\rm Im}(\omega_2/ \omega_1)>0$.
The standard Hamiltonian of the Calogero-Moser system is $H_2$.
As is claimed above (see \cite{Krichever80,Z19}), the Hamiltonian equations 
\beq\label{ham1}
\p_{t_a}x_i=\frac{\p H_a}{\p p_i}, \quad
\p_{t_a}p_i=-\, \frac{\p H_a}{\p x_i}, \quad a=2,3
\eeq
are equations of motion for poles $x_i$ of elliptic solutions
\beq\label{ham2}
u(x, y,t)=-\sum_{i=1}^N \wp \Bigl (x-x_i(y,t)\Bigr ) +\mbox{const}
\eeq
to the KP equation (\ref{kp1}) as functions of $t_2=y$ and $t_3=t$. 
For the BKP equation, the $t_2$-flow is frozen and we will restrict the 
$t_3$-dynamics to the subspace of the phase space with $x_{2i-1}=x_{2i}$, $i=1, \ldots , n$.

\section{Equations of motion for pairs of Calogero-Moser particles}

Let the number of Calogero-Moser particles $N$ be even, $N=2n$. We assume that
the particles are joined in $n$ pairs, so that the coordinates of two particles in each pair
tend to each other:
\beq\label{e1}
x_{2i}-x_{2i-1}=\varepsilon \delta_i, \quad \varepsilon \to 0, \quad i=1,\ldots , n.
\eeq
Here $\delta_i =O(\varepsilon^0)$ as $\varepsilon \to 0$.
We define a submanifold ${\cal B}_n(\varepsilon )={\cal B}_n(\varepsilon ,
\{\delta_i\})$ in the $2N=4n$ dimensional phase space of the 
Calogero-Moser model by imposing the following conditions on the momenta:
\beq\label{e2}
\begin{array}{l}
\displaystyle{
p_{2i-1}=\frac{1}{\varepsilon \delta_i}+\alpha_i \varepsilon +\beta_i\varepsilon^2},
\\ \\
\displaystyle{
p_{2i}=-\frac{1}{\varepsilon \delta_i}-\alpha_i \varepsilon +\beta_i\varepsilon^2}
\end{array}
\eeq
(together with the conditions (\ref{e1})), where
$$
\alpha_i =\alpha_i(\varepsilon )=\alpha_{i,0}+\alpha_{i,1}\varepsilon +
\alpha_{i,2}\varepsilon^2 +\ldots ,
\quad
\beta_i =\beta_i(\varepsilon )=\beta_{i,0}+\beta_{i,1}\varepsilon +
\beta_{i,2}\varepsilon^2 +\ldots
$$
are some series in $\varepsilon$.
Given $\delta_i$, the $\beta_i$ will turn out to be fixed (as we shall see)
and so the 
submanifold ${\cal B}_n(\varepsilon )$ is $2n$-dimensional 
with coordinates $x_{2i-1}, \alpha_i$, $i=1, \ldots , n.$


We are going to show that  
the space ${\cal B}_n(\varepsilon )$ is preserved by the $t_1$-
and $t_3$-flows as $\varepsilon \to 0$ (conjecturally, it is preserved
by all ``odd'' flows). For the $t_1$-flow $\p_{t_1}x_i=-1$, 
$\p_{t_1}p_i=0$ this is obvious.
The $t_3$-flow is given by
\beq\label{e3}
\left \{ \begin{array}{l}
\displaystyle{
\dot x_i=-3p_i^2 +3\sum_{j\neq i}\wp (x_i-x_j),}
\\ \\
\displaystyle{\dot p_i=-3\sum_{j\neq i}(p_i+p_j)\wp '(x_i-x_j),}
\end{array}
\right.
\eeq
where dot means the $t_3$-derivative.
We have by inspection:
\beq\label{e4}
\dot x_{2i-1}=-6\frac{\alpha_i}{\delta_i} +
6 \sum_{j\neq i} \wp (x_{2i-1}-x_{2j-1})+O(\varepsilon )
\eeq
and for consistency we should require that
\beq\label{e5}
\dot x_{2i}-\dot x_{2i-1}=\varepsilon \dot \delta_i,
\eeq
i.e., that evolution in $t_3$ does not generate non-vanishing terms as $\varepsilon \to 0$
(this means that the particles remain to be stuck together).
We have from the first equation in (\ref{e3}):
$$
\begin{array}{lll}
\p_{t_3}(x_{2i}-x_{2i-1})&=&\displaystyle{3(p_{2i-1}^2-p_{2i}^2)}
\\ &&\\
&&\displaystyle{+3\sum_{j\neq i}
\Bigl (\wp (x_{2i}\! - \!x_{2j-1})+\wp (x_{2i}\! - \!x_{2j})\Bigr )}
\\ &&\\
&&-\displaystyle{3\sum_{j\neq i}
\Bigl (\wp (x_{2i-1}\! - \!x_{2j-1})+\wp (x_{2i-1}\! - \!x_{2j})\Bigr )}.
\end{array}
$$
Taking into account that $p_{2i-1}^2-p_{2i}^2=
4\beta_i\delta_i^{-1}\varepsilon +4\alpha_i \beta_i\varepsilon^3$, 
we obtain, expanding the right hand side
in powers of $\varepsilon$, that (\ref{e5}) is true indeed and
\beq\label{e6}
\dot \delta_i=12\beta_{i,0}\delta_i^{-1} +6\delta_i \sum_{j\neq i}\wp '(x_{2i-1}-
x_{2j-1}) \quad \mbox{in order $\varepsilon$},
\eeq
\beq\label{e6a}
4\beta_{i,1}=\delta_i^2\sum_{j\neq i} (\delta_j-\delta_i)\wp ''(x_{2i-1}-x_{2j-1})
\quad \mbox{in order $\varepsilon^2$,}
\eeq
\beq\label{e6c}
24\alpha_i \beta_{i,0}\delta_i +24\beta_{i,2}+\delta_i^2\sum_{j\neq i}
(2\delta_i^2 +3\delta_j^2 -3\delta_i\delta_j)\wp '''(x_{2i-1}-x_{2j-1})=0
\quad \mbox{in order $\varepsilon^3$}.
\eeq
Next, we write the second equation in (\ref{e3}):
$$
\begin{array}{c}
\displaystyle{\dot p_{2i-1}=-3(p_{2i-1}+p_{2i})\wp ' (x_{2i-1}-x_{2i})}
\\ \\
\displaystyle{-3\sum_{j\neq i}\Bigl ((p_{2i-1}+p_{2j-1})\wp '(x_{2i-1}-x_{2j-1})+
(p_{2i-1}+p_{2j})\wp '(x_{2i-1}-x_{2j})\Bigr )}.
\end{array}
$$
Expanding the right hand side in powers of $\varepsilon$ and comparing with (\ref{e2}),
we obtain, in the orders $\varepsilon^{-1}$, $\varepsilon^0$ the same equations (\ref{e6}),
(\ref{e6a}) 
(this shows that the limiting procedure is consistent) and, in the next order,
the equation
\beq\label{e6b}
\dot \alpha_i=-6\alpha_i \sum_{j\neq i}\wp '(x_{2i-1}\! -\! x_{2j-1})-
\frac{3}{2}\sum_{j\neq i}(\delta_i^{-1}\! -\! \delta_j^{-1})\delta_j^2
\wp '''(x_{2i-1}\! -\! x_{2j-1})
-12\delta_i^{-3}\beta_{i,2}.
\eeq

The consistency with the $t_3$-evolution also implies that it should hold
\beq\label{e7}
\dot p_{2i-1}+\dot p_{2i}=O(\varepsilon^2).
\eeq
We have from the second equation in (\ref{e3}):
$$
\begin{array}{lll}
-(\dot p_{2i-1}+\dot p_{2i})&=&
\displaystyle{3p_{2i-1}\sum_{j\neq i}\Bigl (\wp '(x_{2i-1}-x_{2j-1})+
\wp '(x_{2i-1}-x_{2j})\Bigr )}
\\ &&\\
&&\displaystyle{+3p_{2i}\sum_{j\neq i}\Bigl (\wp '(x_{2i}-x_{2j-1})+
\wp '(x_{2i}-x_{2j})\Bigr )}
\\ &&\\
&&\displaystyle{+3\sum_{j\neq i}p_{2j-1}\wp '(x_{2i-1}-x_{2j-1})+3
\sum_{j\neq i}p_{2j}\wp '(x_{2i-1}-x_{2j})}
\\ &&\\
&&\displaystyle{+3\sum_{j\neq i}p_{2j-1}\wp '(x_{2i}-x_{2j-1})+3
\sum_{j\neq i}p_{2j}\wp '(x_{2i}-x_{2j})}.
\end{array}
$$
On the first glance, the right hand side is $O(1)$. However, an accurate 
$\varepsilon$-expansion shows that the terms $O(1)$ and $O(\varepsilon )$ cancel and 
(\ref{e7})
indeed holds. 

It turns out that the $t_3$-flow can be restricted to the submanifold
${\cal B}_n=\lim\limits_{\varepsilon \to 0}{\cal B}_n (\varepsilon )$.
In this way, one can
derive equations of motion for $x_{2i-1}$, 
i.e. the equations connecting $\ddot x_{2i-1}$, $\dot x_{2i-1}$
and $x_{2i-1}$. We have from (\ref{e4}):
\beq\label{e8}
\ddot x_{2i-1}=-6\delta_{i}^{-1}\dot \alpha_i +6\alpha_i\delta_i^{-2}\dot \delta_i +
6\sum_{j\neq i}(\dot x_{2i-1}-\dot x_{2j-1})\wp '( x_{2i-1}-x_{2j-1}).
\eeq
Plugging here $\dot \alpha_i$ from (\ref{e6b}) with $\beta_{i,2}$ from (\ref{e6c}),
we see that all $\delta_i$ cancel and the following equations of motion for $x_i$ hold:
\beq\label{e9}
\ddot x_i+6\sum_{j\neq i}(\dot x_i+\dot x_j)\wp ' (x_i-x_j)-72
\sum_{j\neq i}\sum_{k\neq i} \wp (x_i-x_j)\wp '(x_i-x_k)+6
\sum_{j\neq i} \wp '''(x_i-x_j)=0
\eeq
(here $i,j,k$ are odd numbers running from $1$ to $2n-1$). Using the identity
$\wp '''(x)=12\wp (x)\wp '(x)$, we represent them in the form
\beq\label{e10}
\ddot x_i+6\sum_{j\neq i}(\dot x_i+\dot x_j)\wp ' (x_i-x_j)-72
\sum_{j\neq k\neq i} \wp (x_i-x_j)\wp '(x_i-x_k)=0.
\eeq
These are equations obtained in 
\cite{RZ19} (see also \cite{Z19})
for dynamics of poles of elliptic solutions to the BKP equation. 

In fact the calculations are considerably simplified if one puts $\delta_i=1$ from
the very beginning, i.e. $x_{2i}\! -\! x_{2i-1}=\varepsilon$. 
This fixes the ``gauge freedom'' in the definition
of ${\cal B}_n(\varepsilon )$. The coefficients $\beta_{i,a}$ become functions 
of $x_{2j-1}$, $\alpha_j$ which are independent coordinates in the subspace
${\cal B}_n$. The formulas given above 
show that the restriction of the Calogero-Moser dynamics in $t_3$ (\ref{e3}) 
to ${\cal B}_n$ can be written as a first order system of equations
\beq\label{e11}
\left \{
\begin{array}{l}
\displaystyle{\dot x_{2i-1}=-6\alpha_i +6\sum_{j\neq i}\wp (x_{2i-1}-x_{2j-1}),}
\\ \\
\displaystyle{
\dot \alpha_i = -12\alpha_i \sum_{j\neq i}\wp '(x_{2i-1}-x_{2j-1})+
\sum_{j\neq i}\wp '''(x_{2i-1}-x_{2j-1})}.
\end{array}
\right.
\eeq
Excluding $\alpha_i$, one gets equations (\ref{e10}).

\section{The Lax matrix}

Let us introduce the function
$\Phi (x, \lambda )$ defined as
$$
\Phi (x, \lambda )=\frac{\sigma (x+\lambda )}{\sigma (\lambda )\sigma (x)}\,
e^{-\zeta (\lambda )x},
$$
where $\zeta (\lambda)$ is the Weierstrass $\zeta$-function, i.e. the odd function
such that $\zeta '=-\wp$. 
It has a simple pole
at $x=0$ with residue $1$ and the expansion
$$
\begin{array}{c}
\Phi (x, \lambda )=x^{-1}-\frac{1}{2} \, \wp (\lambda )x 
-\frac{1}{6} \, \wp '(\lambda )x^2 +O(x^3) \quad
\mbox{as $x\to 0$}.
\end{array}
$$
The function $\Phi$ has the following quasiperiodicity properties:
$$
\Phi (x+2\omega_{\alpha} , \lambda )=e^{2(\zeta (\omega_{\alpha} )\lambda -
\zeta (\lambda )\omega_{\alpha} )}
\Phi (x, \lambda ), \quad \alpha =1,2.
$$
In what follows we will often suppress the second argument of $\Phi$ writing simply
$\Phi (x)=\Phi (x, \lambda )$.
We will also need the $x$-derivatives
$\Phi '(x, \lambda )=\p_x \Phi (x, \lambda )$, $\Phi ''(x, \lambda )=\p^2_x \Phi (x, \lambda )$
and so on.

The Lax matrix of the elliptic Calogero-Moser system reads
\beq\label{lax1}
L_{jk}^{\rm CM}=L_{jk}^{\rm CM}(\lambda )
=p_j\delta_{jk}+(1-\delta_{jk})\Phi (x_j-x_k, \lambda ).
\eeq
It is an $N\times N$ matrix depnding on the spectral parameter $\lambda$. 
We are going to restrict it to ${\cal B}_n(\varepsilon )$ as $\varepsilon \to 0$
(with $\delta_i=1$). It is convenient to re-denote $\alpha_i \to \alpha_{2i-1}$. 
The result is a $2n\times 2n$ matrix which is represented as a block matrix with
$2\times 2$ blocks numbered by odd numbers $j,k$ running from $1$ to $2n-1$:
\beq\label{lax2}
\begin{array}{l}
L_{jk}^{(\varepsilon )}=\varepsilon^{-1}\bar e \delta_{jk}+(1-\delta_{jk})
e\Phi (x_j-x_k)
\\ \\
\hspace{2cm}+\varepsilon \Bigl [ (\alpha_j \sigma_3 +\frac{i}{2}\, \wp (\lambda )
\sigma_2 )\delta_{jk} -i(1-\delta_{jk}) \sigma_2 \Phi '(x_j-x_k)\Bigr ] +O(\varepsilon^2),
\end{array}
\eeq
where the matrices $e, \bar e$ are
$$
e=\sigma_0+\sigma_1=\left (\begin{array}{cc}1&1\\1&1\end{array}\right ),
\quad
\bar e=\sigma_3-i\sigma_2=\left (\begin{array}{cc}1&-1\\1&-1\end{array}\right )
$$
and $\sigma_0, \sigma_1, \sigma_2, \sigma_3$ are the standard Pauli matrices
($\sigma_0$ is the unity matrix). For example ($n=2$):
$$
L^{(\varepsilon )}(\lambda )=\left (\begin{array}{cccc}
\varepsilon^{-1}\!+\!\alpha_1\varepsilon & 
-\varepsilon^{-1}\!+\!\frac{1}{2}\, \wp (\lambda )\varepsilon &
\Phi (x_{13}) & \Phi (x_{13})\!-\!\varepsilon \Phi '(x_{13})
\\ &&&\\
\varepsilon^{-1}\! -\! \frac{1}{2}\, \wp (\lambda )\varepsilon &
-\varepsilon^{-1}\! -\! \alpha_1\varepsilon & 
\Phi (x_{13})\! +\! \varepsilon \Phi '(x_{13})&
\Phi (x_{13})
\\ &&&\\
\Phi (x_{31})& \Phi (x_{31})\! -\! \varepsilon \Phi '(x_{31})&
\varepsilon^{-1}\!  +\! \alpha_3\varepsilon &
-\varepsilon^{-1}\! +\! \frac{1}{2}\, \wp (\lambda )\varepsilon 
\\ &&&\\
\Phi (x_{31})\! +\! \varepsilon \Phi '(x_{31})&
\Phi (x_{31})&
\varepsilon^{-1}\! -\! \frac{1}{2}\, \wp (\lambda )\varepsilon &
-\varepsilon^{-1}\! -\! \alpha_3\varepsilon 
\end{array}
\right )
$$
$$
+\, \, O(\varepsilon^2),
$$
where $x_{jk}=x_j-x_k$.
It can be seen that
\beq\label{lax3}
\lim_{\varepsilon \to 0}\, \det_{2n\times 2n} \Bigl (L^{(\varepsilon )}(\lambda )-zI\Bigr )
=R(z, \lambda )
\eeq
exists and $\det\limits_{2n\times 2n} \Bigl (L^{(\varepsilon )}(\lambda )-zI\Bigr )$
is $O(1)$ as $\varepsilon \to 0$. Here $I$ is the unity matrix.
The equation of the spectral curve is 
$R(z, \lambda )=0$. Presumably, this is the same equation as
\beq\label{lax4}
\det_{n\times n}{\cal L}(z, \lambda )=0,
\eeq
where ${\cal L}(z, \lambda )$ is the $n\times n$ 
Lax matrix for pole dynamics of elliptic
solutions to the BKP equation given by \cite{RZ19}
\beq\label{lax5}
\begin{array}{l}
\displaystyle{
{\cal L}_{jk}(z, \lambda )=\Bigl (-\dot x_j +6\sum_{l\neq j}\wp (x_{jk})
-3(z^2 -\wp (\lambda ))\Bigr )\delta_{jk}}
\\ \\
\phantom{aaaaaaaaaaa}-6(1-\delta_{jk})\Phi '(x_{jk}))-6z(1-\delta_{jk})\Phi (x_{jk}).
\end{array}
\eeq
However, we were able to check this explicitly only in the case $n=1$. 
The equations of motion (\ref{e10}) are equivalent to the commutation representation
for the matrix ${\cal L}(z, \lambda )$ in the form of the Manakov's triple \cite{Manakov}:
\beq\label{lax6}
\dot {\cal L}+[{\cal L}, {\cal M}]={\cal P}{\cal L},
\eeq
with some matrix ${\cal M}={\cal M}(z, \lambda )$ depending on the 
dynamical variables, where ${\cal P}$ is a traceless matrix.

\section{Concluding remarks}

We have shown that the third Hamiltonian flow of the elliptic
Calogero-Moser model preserves the configuration in which the particles
join in pairs, with two particles in each pair being placed in one and the same point.
We have also derived the effective equations of motion for the pairs and have shown
that they coincide with equations of motion for poles of elliptic solutions to the 
BKP equation. A problem for the future is to find out whether these equations 
of motion are Hamiltonian or not. 

An advanced approach for analyzing collisions of Calogero-Moser 
particles (in the model with rational potential)
was developed in \cite{KKS78,W98}. It essentially consists in certain completion 
of the phase space which allows one to describe configurations in which coordinates 
of some particles coincide. It would be very interesting to establish connections 
with the approach of that papers, at least in the rational case, where it is available.

\section*{Acknowledgments}

The author thanks I. Krichever and D. Rudneva for discussions. 
The work was performed at the Steklov Mathematical Institute of 
Russian Academy of Sciences, Moscow, and was supported by the 
Russian Science Foundation under grant 19-11-00062.

\end{document}